\documentclass[5p]{elsarticle}
\usepackage{hyperref}
\usepackage{url}
\usepackage{pstricks}
\usepackage{psfrag}
\usepackage{graphicx}
\usepackage{xspace}
\usepackage{multirow}
\usepackage{dcolumn} %centers to the decimal point in tables using ``d''
\usepackage{amssymb}
\usepackage{bm}
\usepackage{scrextend}
\usepackage{tabularx}
\usepackage{amsmath}
\usepackage[T1]{fontenc}
\newcommand{\mevu}{MeV/nucleon}
\newcommand{\ts}{\textsuperscript}
\newcommand{\gammaray}{\ensuremath{\gamma}-ray\xspace}
\newcommand{\gammarays}{\ensuremath{\gamma}-rays\xspace}

\newcommand{\etwop}{\ensuremath{E(2^+_1)}\xspace}
\newcommand{\efourp}{\ensuremath{E(4^+_1)}\xspace}

\newcommand{\etwopt}{\ensuremath{2^+_1 \rightarrow 0^{+}_{\mathrm{gs}}}\xspace}
\newcommand{\efourpt}{\ensuremath{4^+_1 \rightarrow 2^+_1}\xspace}

\newcommand{\stwop}{\ensuremath{2^+_1}\xspace}
\newcommand{\sfourp}{\ensuremath{4^+_1}\xspace}

\newcommand{\bedown}{\ensuremath{{B(E2)\!\!\downarrow}}\xspace}  % short B(E2)

\journal{Physics Letters B}

\makeatletter
\def\ps@pprintTitle{%
 \let\@oddhead\@empty
 \let\@evenhead\@empty
 \def\@oddfoot{}%
 \let\@evenfoot\@oddfoot}
\makeatother

\begin{document}

\begin{frontmatter}

%Title of paper
  \title{Shell evolution of $N=40$ isotones towards \ts{60}Ca: First spectroscopy of \ts{62}Ti}

%% or include affiliations in footnotes:
\author[ariken,alegnaro]{M.~L.~Cort\'es\corref{mycorrespondingauthor}}
\cortext[mycorrespondingauthor]{Corresponding author}
\ead{liliana.cortes@lnl.infn.it}
\author[aunal,ariken]{W.~Rodriguez}
%then in alphabetical order
\author[ariken]{P.~Doornenbal}
\author[acea,atudarmstadt]{A.~Obertelli} 
\author[atriumph]{J.~D.~Holt}   
\author[apadova]{S.~M. Lenzi}  
\author[acns]{J.~Men\'endez}
\author[aiphc]{F.~Nowacki}   
\author[arcnp,aosaka]{K.~Ogata}     
\author[aumadrid]{A.~Poves}     
\author[aumadrid]{T. R. Rodr\'{\i}guez} 
\author[atudarmstadt,aemmi,amaxplankh]{A.  Schwenk}
\author[aprisma]{J. Simonis}
\author[atriumph,awashington]{S.~R.~Stroberg}
\author[ajaea]{K.~Yoshida}
%the rest of the people
%%%spokepersons
%\author{P.~Doornenbal} \ariken
%\author{A.~Obertelli} \atudarmstadt \acea \ariken
%%%%local, MINOS, SAMURAI people, others
\author[acaen]{L.~Achouri}
\author[ariken]{H.~Baba} 
\author[ariken]{F.~Browne}
\author[acea]{D.~Calvet} 
\author[acea]{F.~Ch\^ateau} 
\author[abeijing,ariken]{S.~Chen} 
\author[ariken]{N.~Chiga}
\author[acea]{A.~Corsi}
%\author{M.~L.~Cort\'es} \ariken
\author[acea]{A.~Delbart}
\author[acea]{J-M.~Gheller}
\author[acea]{A.~Giganon}
\author[acea]{A.~Gillibert} 
\author[acea]{C.~Hilaire} 
\author[ariken]{T.~Isobe} 
\author[atohoku]{T.~Kobayashi}
\author[ariken,acns]{Y.~Kubota} 
\author[acea]{V.~Lapoux} 
\author[acea,atudarmstadt,akth]{H.~N.~Liu}
\author[ariken]{T.~Motobayashi} 
\author[aipno,ariken]{I.~Murray} 
\author[ariken]{H.~Otsu} 
\author[ariken]{V.~Panin}
\author[acea]{N.~Paul} 
%\author{W.~Rodriguez}\aunal \ariken
\author[ariken,aut]{H.~Sakurai} 
\author[ariken]{M.~Sasano}
\author[ariken]{D.~Steppenbeck}
\author[acns]{L.~Stuhl}
\author[acea,atudarmstadt]{Y.~L.~Sun}
\author[arikkyo]{Y.~Togano}
\author[ariken]{T.~Uesaka} 
\author[aut]{K.~Wimmer}
\author[ariken]{K.~Yoneda} 
\author[akth]{O.~Aktas}
\author[atudarmstadt,agsi]{T.~Aumann}
\author[ainst]{L.~X.~Chung} 
\author[aipno]{F.~Flavigny}
\author[aipno]{S.~Franchoo}
\author[azagreb,ariken]{I.~Ga\v{s}pari\'{c}}
\author[akoeln]{R.-B.~Gerst}
\author[acaen]{J.~Gibelin}
\author[aewha]{K.~I.~Hahn} 
\author[aewha]{D.~Kim} 
\author[aut]{T.~Koiwai}
\author[atitech]{Y.~Kondo}
\author[atudarmstadt,agsi]{P.~Koseoglou}
\author[ahku]{J.~Lee} 
\author[atudarmstadt]{C.~Lehr}
\author[ainst]{B.~D.~Linh}
\author[ahku]{T.~Lokotko}
\author[aipno]{M.~MacCormick}
\author[akoeln]{K.~Moschner}
\author[atitech]{T.~Nakamura}
\author[aewha]{S.~Y.~Park} 
\author[atudarmstadt]{D.~Rossi}
\author[aoslo]{E.~Sahin} 
\author[aatomki]{D.~Sohler}
\author[atudarmstadt]{P.-A.~S\"oderstr\"om} 
\author[atitech]{S.~Takeuchi}
\author[atudarmstadt,agsi]{H.~Toernqvist}
\author[amadrid]{V.~Vaquero}
\author[atudarmstadt]{V.~Wagner}
\author[alanzhou]{S.~Wang}
\author[atudarmstadt]{V.~Werner}
\author[ahku]{X.~Xu} 
\author[atitech]{H.~Yamada}
\author[alanzhou]{D.~Yan} 
\author[ariken]{Z.~Yang} 
\author[atitech]{M.~Yasuda}
\author[atudarmstadt]{L.~Zanetti}

\address[ariken]   {RIKEN Nishina Center, 2-1 Hirosawa, Wako, Saitama 351-0198, Japan}
\address[alegnaro] {Istituto Nazionale di Fisica Nucleare, Laboratori Nazionali di Legnaro, I-35020 Legnaro, Italy}
\address[aunal]     {Universidad Nacional de Colombia, Sede Bogota, Facultad de Ciencias,\\ Departamento de F\'{\i}sica, Bogot\'a, 111321, Colombia}
\address[acea]      {IRFU, CEA, Universit\'e Paris-Saclay, F-91191 Gif-sur-Yvette, France}
\address[atudarmstadt]{Institut f\"ur Kernphysik, Technische Universit\"at Darmstadt, 64289 Darmstadt, Germany}
\address[atriumph]{TRIUMF, 4004 Wesbrook Mall, Vancouver BC V6T 2A3, Canada}
\address[apadova]{Dipartimento di Fisica e Astronomia, Universit\`a di Padova and INFN, Sezione di Padova, Via F. Marzolo 8, I-35131 Padova, Italy}
\address[acns]      {Center for Nuclear Study, The University of Tokyo, RIKEN campus, Wako, Saitama 351-0198, Japan}
\address[aiphc]       {IPHC, CNRS/IN2P3, Universit\'e de Strasbourg, F-67037 Strasbourg, France}
\address[arcnp]   {Research Center for Nuclear Physics (RCNP), Osaka University, Ibaraki 567-0047, Japan}
\address[aosaka]   {Department of Physics, Osaka City University, Osaka 558-8585, Japan}
\address[aumadrid]{Departamento de F\'{\i}sica Te\'orica and IFT-UAM/CSIC, Universidad Aut\'onoma de Madrid, E-2804 Madrid, Spain}
\address[aemmi]   {ExtreMe Matter Institute EMMI, GSI Helmholtzzentrum f\"ur Schwerionenforschung GmbH, 64291 Darmstadt, Germany}
\address[amaxplankh] {Max-Planck-Institut f\"ur Kernphysik, Saupfercheckweg 1, 69117 Heidelberg Germany}
\address[aprisma] {Institut f\"ur Kernphysik and PRISMA Cluster of Excellence, Johannes Gutenberg-Universit\"at, Mainz 55099, Germany}
\address[awashington] {Department of Physics, University of Washington, Seattle WA, USA}
\address[ajaea]     {Advanced Science Research Center, Japan Atomic Energy Agency, Tokai, Ibaraki 319-1195, Japan}
\address[acaen]    {LPC Caen, ENSICAEN, Université de Caen, CNRS/IN2P3, F-14050 Caen, France}
\address[abeijing]  {State Key Laboratory of Nuclear Physics and Technology, Peking University, Beijing 100871, P.R. China}
\address[atohoku]  {Department of Physics, Tohoku University, Sendai 980-8578, Japan}
\address[akth]      {Department of Physics, Royal Institute of Technology, SE-10691 Stockholm, Sweden}
\address[aipno]    {IPN Orsay, CNRS and Universit\'e Paris Saclay, F-91406 Orsay Cedex, France}
\address[aut]        {Department of Physics, University of Tokyo, 7-3-1 Hongo, Bunkyo, Tokyo 113-0033, Japan}
\address[arikkyo] {Department of Physics, Rikkyo University, 3-34-1 Nishi-Ikebukuro, Toshima, Tokyo 172-8501, Japan}
\address[agsi]       {GSI Helmoltzzentrum f\"ur Schwerionenforschung GmbH, Planckstr. 1, 64291 Darmstadt, Germany}
\address[ainst]     {Institute for Nuclear Science \& Technology, VINATOM, P.O. Box 5T-160, Nghia Do, Hanoi, Vietnam}
\address[azagreb] {Ru{\dj}er Bo\v{s}kovi\'{c} Institute, Bijeni\v{c}ka cesta 54,10000 Zagreb, Croatia}
\address[akoeln]   {Institut f\"ur Kernphysik, Universit\"at zu K\"oln, D-50937 Cologne, Germany}
\address[aewha]   {Department of Science Education and Department of Physics, Ewha Womans University, Seoul 03760, Korea}
\address[atitech]   {Department of Physics, Tokyo Institute of Technology, 2-12-1 O-Okayama, Meguro, Tokyo, 152-8551, Japan}
\address[ahku]     {Department of Physics, The University of Hong Kong, Pokfulam, Hong Kong}
\address[aoslo]     {Department of Physics, University of Oslo, N-0316 Oslo, Norway}
\address[aatomki]  {Institute for Nuclear Research of the Hungarian Academy of Sciences (MTA Atomki), P.O. Box 51, Debrecen H-4001, Hungary}
\address[amadrid] {Instituto de Estructura de la Materia, CSIC, E-28006 Madrid, Spain}
\address[alanzhou]{Institute of Modern Physics, Chinese Academy of Sciences, Lanzhou, China}
\vspace{-1cm}
\begin{abstract}
 Excited states in the $N=40$ isotone \ts{62}Ti were  populated via the \ts{63}V$(p,2p)$\ts{62}Ti reaction at $\sim$200~\mevu~at the Radioactive Isotope Beam Factory and studied using \gammaray spectroscopy.
  The  energies of the \etwopt and \efourpt transitions,  observed here  for the first time,  indicate a deformed \ts{62}Ti ground state.
  These energies are increased compared to the neighboring \ts{64}Cr and \ts{66}Fe isotones, suggesting a small decrease of quadrupole collectivity.
  The present measurement is  well reproduced by large-scale shell-model calculations based on effective interactions, while ab initio and beyond mean-field calculations do not yet reproduce our findings.
  The shell-model calculations for \ts{62}Ti show a dominant configuration with four neutrons excited across the $N=40$ gap.
  Likewise, they indicate that the $N=40$ island of inversion extends down to $Z=20$, disfavoring a possible doubly magic character of the  elusive \ts{60}Ca.
\end{abstract}
\begin{keyword}
  Shell evolution, Radioactive beams, Gamma-ray spectroscopy
%\texttt{elsarticle.cls}\sep \LaTeX\sep Elsevier \sep template
%\MSC[2010] 00-01\sep  99-00
\end{keyword}
\end{frontmatter}

%\linenumbers
Our understanding of atomic nuclei largely  derives from the concept of nuclear shell structure. 
Within this picture, the arrangement of nucleons inside the nucleus can be explained by the filling of discrete  energy levels.
Sizable gaps
between these orbits disfavor the population of the higher-energy levels, and are interpreted as closed shells, which give rise to magic numbers.
%2,8,20,28...
%between these  levels are interpreted as closed shells and give rise to the so-called magic numbers.
Such shell closures can be evidenced by a relatively high-lying first excited 2\ts{+} state,  a  relatively small electric quadrupole transition probability  to the ground state, \bedown, and a steep decrease of the separation energy.
Experimental evidence collected in the last decades, particularly since the advent of radioactive ion beams, has shown that shell structure undergoes significant changes for isotopes far from stability~\cite{Sorlin_PPNP61_2008}.
%magic numbers and how they change
Examples of these changes are the appearance of new magic neutron numbers at $N=32, 34$ in the Ca isotopes and neighboring isotopic chains~\cite{Gade_PRC_74_2006,Wienholtz_Nature498_2013,Steppenbeck_Nature502_2013,Steppenbeck_PRL_114_2015,Rosenbusch_PRL_114_2015,Leistenschneider_PRL_120_2018,Michimasa_PRL_121_2018,Liu_PRL_122_2019},
%although  interpretation is still under debate~\cite{GarciaRuiz_Nature_2016}.
and at  $N=16$ for O isotopes~\cite{Ozawa_PRL_84_2000,Obertelli_PLB_633_2006,Kanungo_PRL102_2009}, 
as well as  the disappearance of the shell closure at $N=8$~\cite{Navin_PRL_85_2000,Iwasaki_PLB_481_2000,Iwasaki_PLB_491_2000, Shimoura_PLB_560_2003}, $N=20$~\cite{Detraz_PRC_19_1979,Motobayashi_PLB_346_1995} and  $N=28$~\cite{Bastin_PRL99_2007,Takeuchi_PRL_109_2012} in various neutron-rich isotopes.

Given that $N=40$, which corresponds to the filling of the neutron $pf$  shells, is a harmonic oscillator  magic number,  the study of the structure of $N=40$ isotones can provide insight into the mechanisms governing shell evolution. 
Indeed the characteristics of this isotonic chain vary with the number of protons.
%Indeed, this isotonic chain shows a variety of characteristics depending on the proton number.
For  \ts{68}Ni ($Z=28$), a  high \etwop energy and a low \bedown  have been observed~\cite{Sorlin_PRL_88_2002}.
%suggesting a magic character of $N=40$.
However, due to the parity change between the $pf$ shell  and the $g_{9/2}$ orbit, the \stwop state   involves at least two neutrons across  $N=40$. Such a neutron-dominated excitation could result in a large \etwop energy and low \bedown value 
without a large shell gap~\cite{Langanke_PRC_67_2003}.
For the  neutron-rich Fe ($Z=26$) and Cr ($Z=24$) isotopes, a monotonous decrease of the \etwop  when approaching $N=40$ and beyond  has been  observed~\cite{Hannawald_PRL_82_1999, Adrich_PRC_77_2008, Santamaria_PRL115_2015, Gade_PRC_81_2010}.
This decrease  indicates a rapid development of collectivity when removing  protons from the $f_{7/2}$ shell.
In contrast, the measurement of the \etwop of \ts{58,60}Ti ($Z=22$) only showed a moderate  decrease  towards $N=40$~\cite{Suzuki_PRC_88_2013,Gade_PRL_112_2014}.
 The very exotic \ts{60}Ca ($Z=20$), where the Ca isotopic chain meets the $N=40$ isotones,  is a  key nucleus for shell evolution~\cite{LRP_EU, LRP_USA}, but  difficult to reach experimentally.
Only recently its  existence  has been established~\cite{Tarasov_PRL_121_2018},
supporting theoretical  predictions for a bound  \ts{70}Ca.
However, the heaviest Ca isotope with known spectroscopic information is  \ts{54}Ca~\cite{Steppenbeck_Nature502_2013}.

Theoretical calculations in the shell-model framework~\cite{Lenzi_PRC_82_2010}  concluded that the development of collectivity in $N = 40$ nuclei is due to quadrupole correlations that give rise to deformed ground states, dominated by intruder neutron orbits beyond the $pf$ shell.
This leads to an island of inversion below \ts{68}Ni, similar to the one formed around \ts{32}Mg~\cite{Lenzi_PRC_82_2010}.
These calculations predict an increase in the \etwop energy of the more exotic $N=40$ isotones  $^{62}$Ti and $^{60}$Ca, while conserving the intruder character in the ground state.
On the other hand, symmetry conserving configuration mixing calculations  with the Gogny interaction predict a conservation of the $N=40$ gap~\cite{Rodriguez_PRC_93_2016}. 
These results agree with calculations performed using
the five-dimension collective Hamiltonian,
which suggest  an energy gap of about 4~MeV at $N=40$, predicting spherical  \ts{62}Ti and \ts{60}Ca~\cite{Gaudefroy_PRC_80_2009, Peru_EurPhysJA_50_2014}.
It is noted that the  beyond-mean-field and the shell model calculations provide similar results for \ts{64}Cr and \ts{66}Fe, while they substantially diverge for \ts{60}Ca and \ts{62}Ti.  Therefore, spectroscopy of \ts{62}Ti offers a crucial test between the two different pictures.
In addition, the properties of Ca isotopes have been extensively studied with coupled-cluster theory~\cite{Hagen_PRL_109_2012} and valence-shell interactions~\cite{Wienholtz_Nature498_2013,Holt_PRC_90_2014}, in both cases using two-nucleon (NN) and three-nucleon (3N) interactions from chiral effective field theory.
Such calculations agree well with experimental energy levels and binding energies up to \ts{54}Ca, and predict the drip line to be located around \ts{60}Ca.
This is in contrast to density functional theories based on the mean field approach which predict, depending on the selected interaction,  Ca isotopes to be bound up to $A=68-76$. 
%Such calculations provide a good agreement with the experimental energy levels and transition probabilities up to \ts{54}Ca.
Beyond $N=40$, coupled-cluster theory  suggests the existence of two-neutron halos and Efimov states in \ts{62}Ca~\cite{Hagen_PRL_111_2013}.

Clearly, spectroscopic information on exotic isotopes around \ts{60}Ca is necessary to deepen our understanding of the nuclear  structure at $N = 40$ and to benchmark the  theoretical predictions towards the neutron drip line.
In the present work, the first spectroscopy of \ts{62}Ti is presented. This isotope represents the closest nucleus to \ts{60}Ca for which spectroscopic studies can be performed at existing radioactive beam facilities.

\begin{figure}[b!]
  \centering
 \includegraphics[width=0.45\textwidth]{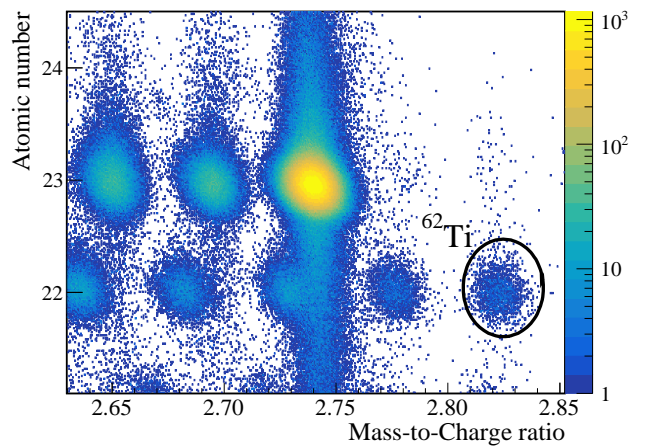}%
 \caption{Particle identification plot  for the outgoing fragments measured with the SAMURAI dipole magnet and related detectors. Incoming \ts{63}V isotopes were selected with BigRIPS. \ts{62}Ti isotopes are indicated by the ellipse.  \label{fig:sampid}}
\end{figure}

The experiment was carried out  at the Radioactive Isotope Beam Factory, operated by the RIKEN Nishina Center
and the Center for Nuclear Study  of the University of Tokyo.
A primary beam of \ts{70}Zn with an energy of 345~\mevu~ and  an average intensity of 240~pnA was fragmented on a 3-mm thick~Be target to produce
a cocktail of secondary beams which included \ts{63}V.
The fragments of interest were selected with the $B\rho - \Delta E - B\rho$ technique using two wedge-shaped aluminium
degraders situated at the dispersive focal planes of BigRIPS~\cite{Kubo_PTEP2012_2012}.
Event-by-event identification was performed by an energy loss measurement  in an ionization chamber,  position and angle measurements 
in  parallel plate avalanche counters at different focal planes, and the time-of-flight measured between two plastic scintillators.
The \ts{63}V isotopes~were delivered 
to the focus area in front of the SAMURAI dipole magnet~\cite{Kobayashi_NIM_317_2013}, with an average intensity of 3~pps and an average energy of 239~\mevu.
At this location the MINOS device~\cite{Obertelli_EPJA50_2014}, composed of a 151.3(13)~mm long liquid hydrogen target surrounded by a Time Projection Chamber (TPC), was placed.
The efficiency of MINOS to detect at least one proton  was measured as 93(4)\% and the resolution for the vertex  reconstruction was estimated to be better than  2~mm ($\sigma$)~\cite{Santamaria_NIMA_905_2018}.
Following proton knockout reactions in the liquid hydrogen target, the \ts{62}Ti fragments had an average energy of 154~\mevu~and  were identified using
the SAMURAI dipole magnet and associated detectors~\cite{Kobayashi_NIM_317_2013}.
%The magnetic rigidity of the fragments was reconstructed based on the position and angle measured in two drift chambers placed before and after SAMURAI.
%The velocity was calculated based on the time-of-flight between a plastic scintillator placed  before the target and a hodoscope located  behind SAMURAI. The hodoscope was also used to obtain the atomic number of the outgoing fragments via an energy loss measurement.
Figure~\ref{fig:sampid} shows the particle identification obtained with SAMURAI when selecting \ts{63}V as incoming beam.  A total of 1880 events corresponding to the \ts{63}V($p,2p$)\ts{62}Ti  reaction was reconstructed.
The transmission of the unreacted \ts{63}V beam along the beam line was measured to be 50.9(11)\% and the inclusive $(p,2p)$ cross section was determined to be 4.0(1)~mb. 

MINOS was surrounded by the high-efficiency $\gamma$-ray detector  array DALI2\ts{+}, composed of 226 NaI(Tl) detectors covering angles between $\sim$15\ts{$\circ$} and $\sim$118$^{\circ}$
with respect to the center of the target~\cite{Takeuchi_NIMA763_2014,Murray_RAPR_2018}.  The array was energy calibrated  using standard  \ts{60}Co, \ts{88}Y, \ts{133}Ba, and \ts{137}Cs sources. 
The full-energy-peak efficiency of the array was determined using a detailed GEANT4~\cite{Agostinelli_NIMA506_2003} simulation and was found to be 30\% at
1~MeV with an energy resolution of 11\% for  a   source moving at 0.6c.
%It was necessary to use a simulation for determining the efficiency of the array due to the extended size of the target.
%Previously reported  efficiency values were in agreement ($\le 6 \%$ error) with the simulation~\cite{Doornenbal_PRC_93_2016}.

\begin{figure}[bt]
  \centering
 \includegraphics[width=0.49\textwidth]{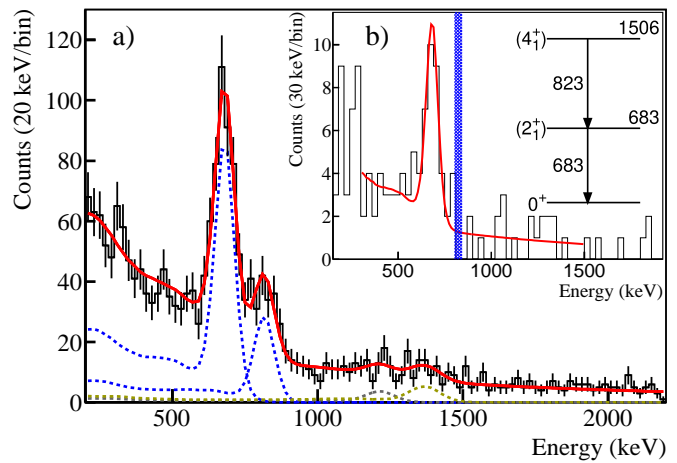}%
 \caption{ a) Doppler corrected \gammaray spectrum of \ts{62}Ti obtained from proton knockout from \ts{63}V. The spectrum was fitted by the convolution of the  simulated  response of DALI2\ts{+} to the observed transitions and a double exponential background.  Two additional transitions are included to improve the fit (see text for details). b) Coincidence spectrum obtained when applying the gate indicated by the  blue area.  \label{fig:spectrum}}
\end{figure}

Doppler corrected $\gamma$-ray spectra were obtained using the reaction vertex and the velocity of the fragment reconstructed with  MINOS.
Peak-to-total ratio and  detection efficiency  improved by adding-up the energies of \gammarays deposited in  detectors up to 10~cm apart. 
To avoid the reconstruction of add-back events from the large atomic background,  $\gamma$-rays with energies below 100~keV were not taken into account in the analysis. 
The Doppler corrected spectrum obtained for the \ts{63}V($p,2p$)\ts{62}Ti reaction is displayed in Fig.~\ref{fig:spectrum}a).
Two peaks are clearly visible  and the $\gamma-\gamma$ coincidence analysis
demonstrates their coincidence (Fig.~\ref{fig:spectrum}b).
%as can be seen in  
Using a 2-dimensional $\chi^2$ minimization, the energies of the transitions were deduced to be 683(10)~keV and 823(20)~keV.
In this minimization procedure, the simulated response of DALI2\ts{+} to transitions of different energies were fitted in steps of 5~keV  to the
experimental data and the $\chi^2$ value was obtained for each combination of energies.
The simulation included the experimental resolution of each crystal and a double exponential background was assumed for the fit.
The parameters of these exponential functions were chosen based on a consistent analysis of the  spectra of proton knockout reactions producing  \ts{50}Ar and \ts{60}Ti. 
The errors on the transition energies include the statistical error from the fit, as well as the systematic error arising from the calibration of the \gammaray detectors 
and the possible lifetime of the states. 
Given that global systematic fits~\cite{Raman_ADNT_78_2001} suggest a lifetime of the \stwop state below 30~ps,  an uncertainty of $15\pm15$~ps was considered for the decay of the \stwop, while the \sfourp was considered short lived.
The best total fit as well as the individual response functions of DALI2\ts{+} are  shown in Fig.~\ref{fig:spectrum}.
The relative intensities of the peaks suggest the tentative  assignment of the 683(10)~keV and the 823(20)~keV peaks  to the  \etwopt and \efourpt transitions, respectively.

A structure  in the \gammaray spectrum above the estimated background was observed between 1000 and 1500~keV.
Two additional transitions at energies of 1222(37)~keV and 1328(45)~keV,  were used to reproduce this structure.
The significance levels of these peaks are $2\sigma$ and $3\sigma$, respectively.
The inclusion of more transitions did not provide any further improvement on the $\chi^2$ of the fit.
%editing referee comment
A structure at 320 keV was  observed with a significance level of 1$\sigma$. The existence of this peak could not be firmly established, therefore  it  was not considered, and its possible contribution to the partial cross section was  assumed to be within the error bars of the analysis.
%end
These possible transitions  indicate the presence of different states being populated in the reaction, but the limited resolution of DALI2\ts{+} and the low statistics did not allow to identify them nor to perform a coincidence analysis.
The existence of such transitions, which  potentially  feed the \stwop or \sfourp states,  implies  a fragmented spectroscopic strength.

Exclusive cross sections to populate the (\stwop) and (\sfourp) states, from which additional feeding should be subtracted,  were calculated based on the fitted \gammaray intensities, the total transmission of the isotopes  and the efficiency of MINOS. Cross sections of 1.5(3)~mb and  0.8(1)~mb were  obtained for the (\stwop) state and the  (\sfourp) state, respectively.
The cross sections measured for the possible transitions at 1222(37)~keV and 1328(45)~keV were determined to be 0.2(1)~mb and 0.3(1)~mb, respectively.
As no firm statement can be made regarding these transitions, we limit the interpretation to their possible direct feeding to the \stwop state. For this, the average value between 100\% feeding and no feeding was considered and the error increased to cover both possibilities, giving a exclusive cross section  of 1.3(4)~mb for the (\stwop) state.

\begin{figure}[t!]
  \centering
 \includegraphics[angle=-90,width=0.5\textwidth]{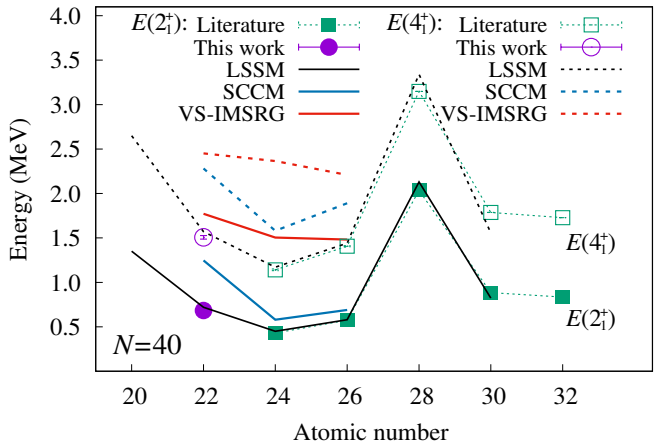}%
 \caption{Systematics of \etwop (filled symbols) and \efourp (open symbols) for even-even $N=40$ isotones. The circles represent the present measurement. The black, blue, and red  lines represent LSSM, SCCM, and VS-IMSRG  calculations, respectively (see text for details).   \label{fig:systematics}}
\end{figure}

The evolution of  measured  \etwop and \efourp  energies for the  $N=40$ isotones between Ti and Ge~\cite{NNDC}  is presented in Fig.~\ref{fig:systematics}.
The  \etwop and \efourp   reported in this Letter for  \ts{62}Ti  have  a similar value than the ones measured for \ts{66}Fe, higher than those of \ts{64}Cr.
It is pointed out that \ts{64}Cr, with a \etwop of 420~keV, has  the largest quadrupole deformation observed in the region~\cite{Gade_PRC_81_2010,Crawford_PRL_110_2013}.
Our results show the  first increase of \etwop along the $N=40$ isotones towards \ts{60}Ca.
This   increase   establishes  a parabolic trend  and suggests a decrease in  quadrupole collectivity.  % towards \ts{60}Ca.
This, in turn,  could be interpreted as a sign of a significant $N=40$ shell gap, and gives  the possibility of a doubly magic character for  \ts{60}Ca.

Large Scale Shell Model (LSSM) calculations, shown by the black lines in Fig.~\ref{fig:systematics}, were carried out with the LNPS interaction~\cite{Lenzi_PRC_82_2010}  using  a \ts{48}Ca core and a  valence space which included the full $pf$ shell for protons and the
$0f_{5/2}$, $1p_{3/2}$, $1p_{1/2}$, $0g_{9/2}$, and $1d_{5/2}$ orbits for  neutrons. 
This interaction has already successfully reproduced the \etwop of the heavier $N=40$ isotones~\cite{Lenzi_PRC_82_2010}.
%As can be seen from the Figure, t
The  LSSM calculations  reproduce very accurately the data for both the \etwop and \efourp  of the $N=40$ isotones including our values for  \ts{62}Ti. This agreement strengthens the tentative spin and parity assignment for these states.
%The increase of the \etwop and \efourp energies compared with the neighboring \ts{64}Cr can be interpreted  as a decrease in the collectivity of \ts{62}Ti.
As shown in Ref.~\cite{Lenzi_PRC_82_2010},  the calculations predict  a reduction of the $0f_{5/2}-0g_{9/2}$ gap when going from \ts{68}Ni to \ts{60}Ca, as well as the closeness of the quadrupole partner orbits $0g_{9/2}$ and $1d_{5/2}$. 
Due to this proximity, quadrupole correlations  produce a gain in energy that largely overcomes the cost of exciting neutrons across the $N=40$ gap, thereby favoring many-particle-many-hole configurations.
This  situation resembles the behavior at $N=20$ and suggests an island of inversion for $N=40$ isotones below \ts{68}Ni.
For  \ts{62}Ti,  a gap of about 1~MeV is predicted,  with  a  resulting wave function  dominated by 4p-4h excitations (63\%) and a significant 6p-6h  component (22\%)~\cite{Lenzi_PRC_82_2010}. Furthermore, a  ground-state deformation parameter  $\beta=0.28$ for \ts{62}Ti is obtained.
%Most notably, the agreement with the measured energies in \ts{62}Ti
The agreement with the measured energies  of the $N=40$ isotones, including  \ts{62}Ti, 
indicates  that the island of inversion in this region extends down to \ts{60}Ca.
%Most notably,
 It is particularly remarkable that although the \etwop for \ts{60}Ca is predicted to be 1.35~MeV,  which  represents an increase with respect to the neighboring isotones, a  4p-4h configuration dominance  (59\%) prevails~\cite{Lenzi_PRC_82_2010}.

Symmetry conserving configuration mixing (SCCM)  calculations using the  Gogny D1S effective interaction~\cite{Decharge_PRC_21_1980, Berger_NPA_428_1984} were performed 
for \ts{62}Ti, \ts{64}Cr, and \ts{66}Fe, and are indicated by the blue lines in Fig.~\ref{fig:systematics}.  
For the calculations, each individual nuclear state was defined as the linear combination of multiple intrinsic many-body states
with different quadrupole (axial and triaxial) shapes~\cite{Rodriguez_PRC_81_2010,Rodriguez_PRC_93_2016}.
Cranked or octupole deformed states were not included, 
therefore, a systematic stretching of the levels with respect to the experimental values is expected~\cite{Borrajo_PLB_746_2015,Robledo_JPG_46_2018}. 
The \etwop predicted for  \ts{64}Cr and \ts{66}Fe lie very close to the LSSM predictions, and are in fair agreement with the experimental data. However, when going to \ts{62}Ti, a more abrupt increase of the \etwop  is obtained.  For the \efourp energies, the calculations  overestimate the experimental values by about 500~keV, although the minimum value for \ts{64}Cr is maintained. It is noted that for \ts{64}Cr and \ts{66}Fe, where the deformation is well described by the model,  the inclusion of cranking would further improve the agreement with the experimental data. 
Within this model, the energy gap at $N=40$ is conserved, leading to a ground state of \ts{62}Ti  highly mixed with the spherical configuration.
This is also the case for  \ts{60}Ca, which  is predicted as a doubly magic nucleus with an \etwop of 4.73~MeV~\cite{Robledo_JPG_46_2018}. 
%Indeed,  a deformation parameter of $\beta<0.1$ is obtained for \ts{62}Ti, while  \ts{60}Ca is predicted as a doubly magic nucleus.
It is noted that although this calculation yields a spherical ground state for \ts{62}Ti, the \stwop and \sfourp states belong to a deformed band starting at the  $0^+_2$ state. This band can correspond to the predictions of the LSSM calculations and  indicate that the SCCM calculations overestimate the $N=40$ gap in this region.
%would indicate that the SCCM calculation underestimates the  correlations in this region. 

Ab initio valence-space in-medium similarity renormalization group (VS-IMSRG)~\cite{Tsukiyama_PRC_85_2012, Bogner_PRL_113_2014, Stroberg_PRC_93_2016,Stroberg_PRL_118_2017,Stroberg:2019mxo} calculations were also performed for  $^{62}$Ti, $^{64}$Cr, and $^{66}$Fe, as shown by the red lines in Fig.~\ref{fig:systematics}.
The chiral NN+3N interaction labeled 1.8/2.0 (EM) in Refs.~\cite{Simonis_PRC_96_2017,Hebeler_PRC_83_2011} was used, which is based on the NN potential from Ref.~\cite{Entem_PRC_68_2003} and 3N forces fitted to light systems up to \ts{4}He only. With this NN+3N interaction,
%The chiral effective field theory nucleon-nucleon (NN) and three-nucleon (3N) interaction labeled 1.8/2.0 (EM) in Refs.~\cite{Simonis_PRC_96_2017,Hebeler_PRC_83_2011}, based on the  NN interaction from Ref.~\cite{Entem_PRC_68_2003} and 3N forces fitted to light systems up to $^4$He, was used (see  Ref.~\cite{Simonis_PRC_96_2017} for more details).
%With this model,
ground-state energies up to Sn~\cite{Stroberg:2019mxo,Simonis_PRC_96_2017,Simonos_PRC_93_2016, Morris_PRL_120_2018}  are generally well reproduced.
As the VS-IMSRG captures 3N forces between valence nucleons via an ensemble normal ordering ~\cite{Stroberg_PRL_118_2017}, a separate
valence-space interaction is decoupled for each nucleus of interest.
Here, the same model space as the LNPS Hamiltonian is considered (adding the $2s_{1/2}$ neutron orbital for \ts{62}Ti).
Using the Magnus formulation of the IMSRG~\cite{Morris_PRC_92_2015}, operators at the two-body level are truncated in the so-called IMSRG(2) approximation.
%in the same model space as the LNPS Hamiltonian (adding the $2s_{1/2}$ neutron orbital for \ts{62}Ti).
%Within the Magnus formulation of the IMSRG~\cite{Morris_PRC_92_2015} operators are truncated at the two-body level, the so-called IMSRG(2) approximation.
The VS-IMSRG interaction is diagonalized with the code ANTOINE~\cite{Caurier_RevModPhys_77_2005},
including, for the first time in the VS-IMSRG, both intruder quadrupole partners, such as $0g_{9/2}$-- $1d_{5/2}$~\cite{Mougeot_PRL_120_2018}.
%including for the first time in the VS-IMSRG extruder quadrupole partners, such as $0g_{9/2}$-- $1d_{5/2}$, in potentially deformed nuclei.
%Figure ~\ref{fig:systematics}  shows that
The VS-IMSRG overestimates the  \etwop and \efourp excitation energies in
\ts{62}Ti, \ts{64}Cr, and \ts{66}Fe, predicting all states as  spherical.
Cross-shell excitations to the $0g_{9/2}$-- $1d_{5/2}$ orbits stay at the 1p-1h level because of the substantial $N=40$ shell gap, 3.7~MeV in \ts{62}Ti.
Within this model, a  \etwop of around 7~MeV is predicted  for \ts{60}Ca,  an overestimation  which is also observed at other shell closures with the VS-IMSRG~\cite{Simonis_PRC_96_2017, Morris_PRL_120_2018,Taniuchi_Nature_569_2019}.
%also gives \etwop  that are too high at other shell closures
This limitation has been related to the IMSRG(2) truncation~\cite{Mougeot_PRL_120_2018}, which may not fully capture correlations associated with cross-shell excitations.
Preliminary comparisons with coupled-cluster theory indicate that keeping operators at the three-body level will improve the results.
Also, choosing a deformed reference state, instead of spherical as in the present work, may capture quadrupole correlations more efficiently~\cite{Hergert_JoPG_1041_2018,Yao_PRC_98_2018}.

%%%%%%%%%%% edit for the cross sections %%%%%%%%%%%%%%%%%%%%%%%%%%%%%%%%%%%%5

\setlength\extrarowheight{2pt}
\begin{table*}[h!]
  {%\footnotesize
  \caption{Experimentally deduced excitation energies and cross sections for \ts{62}Ti following the \ts{63}V$(p,2p)$\ts{62}Ti reaction, and comparison with theoretical cross sections obtained with the LSSM calculation.  The spectroscopic factors and corresponding cross sections are shown for the three possible values of the spin and parity of the ground state of \ts{63}V. The experimental ground-state cross section was calculated by subtracting the cross sections of the measured transitions  from the inclusive cross section. \label{tab:CrossSections}}
  \begin{tabular}{|cc|cccc|cc|cc|cc|}\hline\hline
    
\multirow{2}{*}{$ E$ (keV)}&\multirow{2}{*}{$\sigma_{\text{exp}}$ (mb)}&\multirow{2}{*}{$E$ (keV)} &\multirow{2}{*}{ J$^\pi$}& \multirow{2}{*}{$l_j $} &   $\sigma_{s.p}$ (mb) &\multicolumn{2}{c|}{$J^\pi=3/2^-$ }&\multicolumn{2}{c|}{$J^\pi=5/2^-$ }&\multicolumn{2}{c|}{$J^\pi=7/2^-$ } \\
                                             &                                                             &                                              &                                          &                                       &                                   &    $C^2S$  &    $\sigma_{\text{theo}}$ (mb) &    $C^2S$  &   $\sigma_{\text{theo}}$ (mb)   & $C^2S$  & $\sigma_{\text{theo}}$ (mb)  \\\hline 
   \multirow{2}{*}{0}                  &   \multirow{2}{*}{1.4(4)}   &    \multirow{2}{*}{0}   &    \multirow{2}{*}{0$^+_1$} &  $p_{3/2}$  &        1.56                        &      0.03     & \multirow{2}{*}{0.05}    &       ---       &  \multirow{2}{*}{0.04}&     ---       & \multirow{2}{*}{0.58} \\
                                                     &                                              &                                       &                                                 &  $f_{7/2}$   &        1.46                        &       ---        &                                         &    0.03      &                                       &   0.4       &                                    \\\hline
 \multirow{2}{*}{683(10)}         &   \multirow{2}{*}{1.3(4)} &    \multirow{2}{*}{720}&   \multirow{2}{*}{2$^+_1$} &  $p_{3/2}$  &        1.54                        &      0.06     & \multirow{2}{*}{0.61}   &    0.01      &  \multirow{2}{*}{0.97} &   0.02     & \multirow{2}{*}{0.07} \\
                                                     &                                              &                                       &                                                 &  $f_{7/2}$   &        1.44                        &       0.36    &                                        &    0.66      &                                        &   0.03     &                                    \\\hline
\multirow{2}{*}{1506(22)}        &   \multirow{2}{*}{0.8(1)}& \multirow{2}{*}{1570}&    \multirow{2}{*}{4$^+_1$} &  $p_{3/2}$  &        1.50                        &      ---        & \multirow{2}{*}{1.30}    &      0.04      &  \multirow{2}{*}{0.38}&  0.04       & \multirow{2}{*}{0.44} \\
                                                     &                                              &                                       &                                                 &  $f_{7/2}$   &        1.41                        &       0.92        &                                     &    0.23      &                                       &   0.27       &                                    \\\hline\hline
  
  \end{tabular}
  }
\end{table*}

\begin{figure}[t!]
 \includegraphics[angle=-90,width=0.49\textwidth]{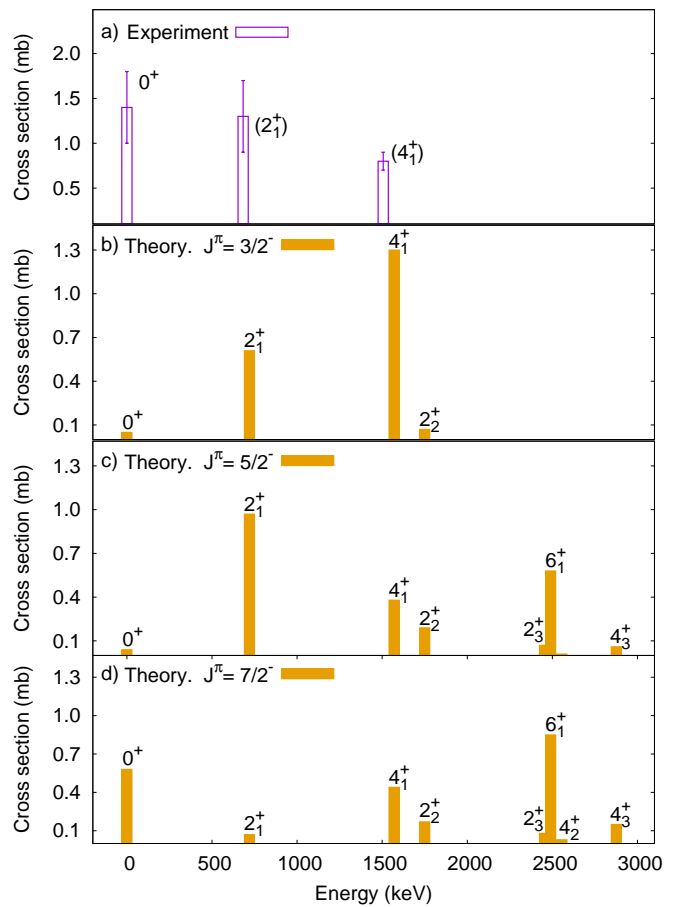}%
 \caption{Partial proton removal cross sections for the \ts{63}V$(p,2p)$\ts{62}Ti reaction. Panel a) shows the experimental results. Panels b) to d) show  LSSM calculations using the LNPS interaction  assuming the ground state of \ts{63}V as $3/2^-$, $5/2^-$ and $7/2^-$, respectively. \label{fig:sigma}}
%   In panel e) the proposed level scheme based on the experimental data is shown. Panel f) displays the level scheme obtained using  LSSM calculations. For each state only the transition with highest strengths is  shown.   
\end{figure}

%Theoretical cross sections were obtained as described in the supplementary material~\cite{supplementary}.
Single-particle theoretical cross sections  were computed in  the  DWIA framework~\cite{Wakasa_PPNP_96_2017}.
The single-particle wave functions and the nuclear density were obtained by the Bohr-Mottelson single-particle potential~\cite{Bohr-Mottelson}.
The optical potentials for the distorted waves in the initial and final channels  were constructed by the microscopic folding model~\cite{Toyokawa_PRC_88_2013} with the Melbourne G-matrix interaction~\cite{Amos_ANP_25_2000} and with the calculated nuclear density. The spin-orbit part of each distorting potential was disregarded.
As for the transition interaction, the Franey-Love effective proton-proton interaction was adopted~\cite{Franey_PRC_31_1985}.
%To compare the calculated  cross sections to the measured ones, the
Cross sections at different beam energies, from 240~\mevu~at the entrance of the target to 154~\mevu~at the exit, were calculated and weighted according to the energy loss in the target.
Theoretical cross sections ($\sigma_{\text{theo}}$) were  obtained by weighting the single particle cross sections by  the calculated spectroscopic factors.

The spin and parity of the  ground state of \ts{63}V are not known experimentally. The LSSM calculation suggests it to be $3/2^-$,  although states with spin and parity of $5/2^-$ and $7/2^-$ appear very close in energy,  suggesting the  presence of isomeric states. No experimental evidence of  such states has been reported so far and available data are consistent with a $3/2^-$ assignment~\cite{Suchyta_PRC_89_2014}.  Results of the calculations for the three cases are shown in Table~\ref{tab:CrossSections}, and displayed in Fig.~\ref{fig:sigma}, together with the experimental results.
%The experimental ground-state cross section was calculated by subtracting the cross sections of the measured transitions  from the inclusive cross section.
%Fig.~\ref{fig:sigma}  shows a comparison between the measured cross sections and theoretical cross sections to populate low-lying states in \ts{62}Ti. The three possible values for the ground-state spin of the incoming  \ts{63}V were considered.
%To better compare  to the experimental result, a reduction factor of 0.6 was applied to the theoretical cross sections~\cite{Atar_PRL_120_2018}. 
It can be seen that  neither the absolute value or the general trend shown by the data are reproduced by the calculation in any scenario. 
The calculation for the ground state  of $J^{\pi}=3/2^-$ resembles better the experimental data in terms of  the number of states that are populated, while for the cases of $J^{\pi}=5/2^-$ and $J^{\pi}=7/2^-$ a considerable population of the $6^+_1$ state would be expected. In particular for the case of $J^{\pi}=7/2^-$ a population of the $6^+_1$ state higher  than the one of the  \stwop state would be expected, at odds with the experimental result. 
It is noted that the calculated spectroscopic factors add up to less than half  of the total strength in the three cases. 
Therefore, population of  higher lying states  is expected  by the calculations.
Such a scenario would lead to  unobserved  transitions feeding the \sfourp  or the \stwop states directly, which can account for the excess of the  measured cross section in comparison with the calculations.
Although not in good agreement, the low measured and calculated partial cross sections, as well as the apparent fragmentation of the spectroscopic  strength, are consistent with the collective nature of the \ts{62}Ti ground state discussed in this work. However, the large error bars prevent a firmer conclusion.

In summary, first spectroscopy of \ts{62}Ti was obtained by means of the  \ts{63}V$(p,2p)$\ts{62}Ti reaction at $\sim$200~\mevu.
Transitions at 683(10)~keV and 823(20)~keV  were assigned to the decay of the \stwop and \sfourp states at 683(10)~keV and 1506(22)~keV, respectively.
Our result shows for the  first time an  increase  of the  \etwop for $N=40$ isotones towards \ts{60}Ca.
 LSSM calculations were in good agreement with the experimental findings. 
 The calculations suggest that although the collectivity decreases approaching \ts{60}Ca, with an ensuing increase of \etwop,  quadrupole correlation contributions remain and lead to
 the extension of the $N=40$ island of inversion down to \ts{60}Ca.
SCCM calculations overestimate the measured \etwop and \efourp of \ts{62}Ti,  
predicting a doubly magic character of \ts{60}Ca and a weakly deformed ground state in \ts{62}Ti, at variance with the LSSM calculations.
For these calculations the $N=40$ spherical gap is too large to produce the inversion between the quasi-spherical and deformed $0^{+}$ states.
VS-IMSRG calculations, which provide a good description of excited states in Ca isotopes,   largely overestimate the \etwop and \efourp energies of \ts{62}Ti, even after the inclusion of the neutron  $0g_{9/2}$, $1d_{5/2}$ and $2s_{1/2}$ orbitals.  The spectroscopic information presented in this Letter offers an important benchmark for our understanding of nuclear structure  approaching \ts{60}Ca and the location of the  neutron drip line. 

%\section*{Acknowledgments}

  We thank the RIKEN Nishina Center accelerator staff and the BigRIPS team for the stable operation of the high-intensity  Zn beam and for the preparation of the secondary beam  setting. 
  K.O. acknowledges the support by Grant-in-Aid for Scientific Research  JP16K05352.
  A.P. is supported in part by the Ministerio de Ciencia, Innovacion y Universidades (Spain), Severo Ochoa Programme SEV-2016-0597 and grant PGC-2018-94583.
  F.B. is supported by the RIKEN Special Postdoctoral Researcher Program.
  L.X.C. and B.D.L would like to thank MOST for its support through the Physics Development Program Grant No.{\DJ}T{\DJ}LCN.25/18.
  I.G. has been supported by HIC for FAIR and Croatian Science Foundation under projects no. 1257 and 7194.
  D.So. was supported by projects No. GINOP-2.3.3-15-2016-00034 and No. K128947.
  V.V. acknowledges support from the Spanish Ministerio de Economía y Competitividad under Contract No. FPA2017-84756-C4-2-P.
  K.I.H., D.K. and S.Y.P. acknowledge the support from the NRF grant funded by the Korea government (No. 2018R1A5A1025563 and 2019M7A1A1033186).
  The development of MINOS was supported by the European Research Council through the ERC Grant No. MINOS-258567.
  This work was also supported by the   NKFIH (128072), the   JSPS KAKENHI Grant No. 18K03639, MEXT as “Priority issue on post-K computer” (Elucidation of the fundamental laws and evolution of the universe), JICFuS, the CNS-RIKEN joint project for large-scale nuclear structure calculations, NSERC, the Deutsche Forschungsgemeinschaft -- Projektnummer 279384907 -- SFB 1245, the PRISMA Cluster of Excellence, and the BMBF under Contracts No.~05P18RDFN1 and 05P19RDFN1. TRIUMF receives funding via a contribution through the National Research Council Canada. Computations were performed at the J\"ulich Supercomputing Center (JURECA).

%\bibliography{References}

\end{document}